\documentclass[12pt]{article}
\usepackage{amsfonts}
\usepackage{amscd}
\usepackage{amsmath}
\begin{document}

\begin{center}
\large{\textbf{ANZATZ OF QUANTUM CORRELATION\\
AND DISENTANGLEMENT PROBLEM}\\}

\smallskip
N. K. Solovarov\\
Zavoisky Physical-Technical Institute of Russian Academy of
Sciences, Kazan, 420029 Russia\\
Email: solovar@kfti.knc.ru\\
\end{center}
\begin{abstract}
Pedagogical introduction into the problem of the mathematical
description of the quantum correlation (entanglement) of composite
quantum systems is represented. The notion is substantiated about
the fact that the conventional algorithm of the reduction of von
Neumann in the description of the dynamics of the observed
subsystem is not universal and corresponds only to the case of
maximum macroscopicity of the unobservable subsystem. Is clearly
shown the sense of the algorithm of the correlated reduction
proposed, which minimally changes the entropy of composite system.
\end{abstract}


 \section{Introduction. Theoretical difficulties in
 the description of quantum correlation.}

The property of quantum correlation (or entanglement) of composite
quantum systems became last decade one of the most discussed
questions of quantum physics. The number of reviews and articles,
in which is used the idea about the quantum correlation,
continuously and almost exponentially grows according to the
Gisin's estimation \cite{gisin} since 1995. The reasons for this
avalanche-type interest several, and even classification reasons
appears ambiguous. Here we first attempt ourselves to trace how
idea about the quantum correlation is expressed mathematically
and, to what experimental operations correspond the accepted
mathematical models . Primary attention is paid to the
difficulties and the contradictions in the description and the
interpretation of the experimental manifestations of quantum
correlation. In the introductory part of the report we refer
mainly to the last reviews, where it is possible to find the
comprehensive bibliography.

The study of quantum correlation is conducted at present from
different positions. Historically idea about the quantum
correlation appeared with the examination of physical nature of
quantum non-locality \cite{gisin}. Another approach is connected
with the description of the decogerence phenomenon of quantum
systems due to the measurement and it is called frequently the
problem of quantum measurements \cite{dass,zeh}. The third, most
"published" (according to the number of articles) approach, is
connected with the development of quantum information theory and
the examination of the possibility of using the quantum
correlation for transfer and processing of information
\cite{batle,plenio}. In spite of the use of different terminology
and interpretation, in all enumerated approaches it is easy to
trace the united mathematical and physical content. Our first
purpose, to show mathematical generality and physical content of
differences and difficulties, which are in the different
approaches.

Let us begin from the history, i.e., let us trace, what debating
point, and what logical sequence of ideas and experiments led to
the present situation.

In the work \cite{gisin} this logical sequence of ideas is erected
on the way of examining the axiomatically adopted (being seemed a
priori obvious) property of the locality of nature - interaction
occurs three-dimensional locally, the transfer of interaction
between the three-dimensional different points is accomplished by
certain material agent, the velocity of propagation of which is
limited. Newton clearly formulated the being contained in its
theory of universal gravitation (but by them categorically
rejected) property of the non-locality (stone moved on the Moon
instantly changes the weight of any object on the Earth). Newton's
non-locality mathematically is expressed in the absence in the law
of universal gravitation of dependence on the time. To this,
emphasized by Newton internal contradiction in the mathematical
description of nature (mathematical model of gravity) the majority
of physicists did not turn attention, since its experimental
manifestations it was not observed. Non-locality (in the theory)
was present up to 1915, when Einstein formulated general theory of
relativity. Einstein by general theory of relativity returned to
physics locality \cite{gisin}.

Quantum mechanics was borne in 10 years and again in the
description of nature non-locality returned. Einstein in principle
did not agree with the statistical interpretation of quantum
mechanics, expressing the sense of disagreement by the known
phrase:
\begin{quote}
\emph{God do not plays dice!}
\end{quote}
Paraphrasing the statement of Einstein Gisin \cite{gisin}
formulates the fundamental results of the development of quantum
physics of last decade by the assertion:
\begin{quote}
\emph{God does play dice, he even plays with nonlocal dice!}
\end{quote}

Where in the quantum theory is contained quantum non-locality
(quantum correlation, entanglement) and how it is expressed
mathematically?

The unavoidable preamble of answer to this question is reminding
the axioms of quantum mechanics and interpretation accepted, (see,
for example, the recent account of this preamble in the thesis
\cite{batle}). Let us limit this introduction, after isolating
only those positions, around which now goes the discussion about
the quantum correlation.

With the description of the quantum correlation:
\begin{enumerate}
\item One of the postulates of quantum theory is used - reduction
postulate, i.e. the action of measurement consist in the collapse
of wave function.
\item The tasks of quantum dynamics of composite
quantum systems are examined.
\end{enumerate}

Let the state of quantum system be represented by the wave
function
$\vert\Psi_{l}\rangle=\sum_{\alpha}(c_{l\alpha}\vert\alpha\rangle)$,
$\sum_{\alpha}\vert c_{l\alpha}\vert^{2} =1$, where
$\vert\alpha\rangle$ is the complete orthonormal set of the
eigenstates (basis) of the considered system, or by the density
operator $\hat\rho=\sum_{l}
p_{l}\vert\Psi_l\rangle\langle\Psi_l\vert$, $\sum_{l}p_l=1$ ,
where the pure states $\vert\Psi_{l}\rangle$ are not compulsorily
orthogonal. The set of eigenstates of the Hamiltonian of physical
system is selected in many instances as the basis.

Let that observable $\hat A$ has eigenvalues $a$ with the
appropriate eigenvectors $\vert ad\rangle$, where the additional
index $d$ noted possible degeneracy $a=ad$. Then projector to the
subspace of that observed $\hat A$ with the eigenvalue $a$ is an
operator $\hat P_{a}=\sum_{ad=a}\vert ad\rangle\langle ad\vert$.
Mathematically the reduction postulate is expressed by the
following conversion of clean wave function as a result of
macroscopic measurement of observable $a$:
\begin{equation}\label{eq1}
  \vert\Psi_{l}\rangle\to\vert ad\rangle=\frac{\hat
  P_{ad}\vert\Psi_{l}\rangle}{\sqrt{\langle\Psi_{l}\vert\hat
  P_{ad}\vert\Psi_{l}\rangle}}.
\end{equation}

Normalizing coefficient in the radicand of denominator is equal to
the probability of the realization of this event. We will note
that in the right side of~(1) the equality is carried out with an
accuracy to the phase factor, which is considered usually
unessential. If degeneration is absent, then this conversion is
considered describing the ideal quantum von Neumann's measurement.

The conversion for the density operator corresponding to reduction
is accepted to write in the form \cite{batle}
\begin{equation}\label{eq2}
  \hat\rho\to\hat\rho_{ad }=\frac{\hat
P_{ad}\hat\rho\hat P_{ad}^{\dag}} {\textrm{Sp}\left(\hat
P_{ad}\hat\rho\hat P_{ad}^{\dag}\right)},
\end{equation}
with the assumptions of i) orthogonality $\textrm{Sp}\left(\hat
P_{ad}\hat P_{ag}^{\dag}\right)=\delta_{dg}$ and ii) closure
$\sum_{ad}\hat P_{ad}\hat P_{ad}^{\dag}=\hat I$, where $\hat I$ is
the unit operator and subscript $\textrm{N}$ here and subsequently
will be used for the designation of that reduction according to
the axiomatic von Neumann's algorithm. Normalizing factor in
denominator of (2) it is possible to copy in the different form,
since the trace does not depend on the non-diagonal matrix
elements of operator. Conversions~(1),(2) relate to the arbitrary
closed quantum system, on which is conducted the measurement by
external macroscopic measuring device. It is necessary to
emphasize that very start of measurement is passage to the open
system, but all the examination is conducted in the Hilbert space
of the closed system.

The property of quantum correlation is determined for the
composite quantum systems. Generally speaking, in quantum physics
the composite systems are always examined, since the reduction
postulate, being the inherent part of the quantum theory, implies
existence of two physically divided parts: the described quantum
system and measuring (macroscopic) device. However, usually,
keeping in mind composite quantum system, it is assumed that they
are determined (i.e. experimentally they can be independently
isolated and studied) several quantum subsystems $A,B,E,\ldots$(in
the quantum informatics corresponding terms are accepted - Alice,
Bob, Eve-Eavesdropper, Charlie...). The certainty of such
subsystems means that are considered known their Hamiltonians
$\hat H_{A},\hat H_{B},\hat H_{E},\ldots$, which correspond
eigenstates
$\vert\alpha\rangle,\vert\beta\rangle,\vert\varepsilon\rangle,\ldots$
(frequently selected as the bases), the operators of those
observable $\hat A,\hat B,\hat E,\ldots$, their eigenvalues
$a,b,e,\ldots$ and eigenstates $\vert a\rangle,\vert
b\rangle,\vert e\rangle,\ldots$. We for the simplification not
will here introduce special indices for the designation of
possible degeneracy. Furthermore, are considered known the
Hamiltonians of pairwise interactions between the subsystems $\hat
H_{AB},\hat H_{AE},\hat H_{BE},\ldots$, moreover interactions are
relied by the relatively weak $\left(\hat H_{AB},\hat
H_{AB},\ldots\right)<\left(\hat H_{A},\hat H_{B},\ldots\right)$
(there are in the form the relative values of the eigenvalues of
energy). The obvious terminology is accepted: bipartite system
$\left(\hat H_{A+B}\right)$, tripartite system $\left(\hat
H_{A+B+E}\right)$,\ldots multipartite system $\left(\hat
H_{A+B+E+\ldots}\right)$.

The majorities of the fundamental properties of quantum
correlation can be examined based on the example of the bipartite
quantum system $\left(\hat H_{A+B}\right)$. The relative smallness
of interactions between the subsystems makes possible to use the
first mathematical assumption - to describe the state of composite
system in the Hilbert space of dimensionality $N_{A}\times N_{B}$,
which is been the direct product space of the space of the
independent subsystems , where $N_{A}, N_{B}$ are the
dimensionality (number of eigenstates) of subsystems respectively.
I.e. the set of pared multiplications of eigenvectors of
subsystems $\vert\alpha\rangle\times\vert\beta\rangle$ is selected
as basis. Let the state of system in this basis be described by
the density operator of  $\hat\rho_{A+B}$ ($\hat\rho_{A+B}\in\hat
H_{A}\otimes\hat H_{B}$).

There is one additional taciturn adopted (implicit) limitation to
many composite quantum systems, for which is determined the
property of quantum correlation. It is assumed that there is a
physical possibility to conduct local actions or measurements on
the subsystems \cite{terhal}. Term "local" here does not bear in
the general case of the content "three-dimensional localization",
although in many experimental cases this precisely thus.
Mathematically this possibility is described by assumption about
the validity of existence of the following (local) maps of the
states of the quantum system:
\begin{equation}\label{eq3}
  \hat\rho'_{A+B}=\left(\hat U_{A}\otimes\hat U_{B}\right)
  \hat\rho_{A+B}\left(\hat U_{A}\otimes\hat U_{B}\right)^{\dagger},
\end{equation}
where $\hat U_{A},\hat U_{B}$ are unitary operators, determined in
the spaces of the subsystems, by which is described the action
(most frequently external) on the subsystems. The equality $\hat
U_{A(B)}=\hat I_{A(B)}$, where $\hat U_{A}$ are unit operators in
the space of the corresponding subsystem, corresponds to the
absence of local action on the subsystem.

Existence of quantum correlation (entanglement) between the
subsystems of such composite system is accepted to mathematically
determine through the opposite property of the inseparability of
quantum system, i.e., the impossibility to represent the density
operator in the form of the convex linear combination of the
direct (tensor) products of the pure density operators of
subsystems \cite{werner}:
\begin{equation}\label{eq4}
  \hat\rho_{(A+B)s}=\sum_{l}
p_{l}\left(\hat\rho_{Al}\otimes\hat\rho_{Bl}\right),
\end{equation}
 where
$0<p_{l}\le 1$, $\sum_{l} p_{l}=1$,
$\left(\hat\rho_{Al}\right)^2=\hat\rho_{Al}$,
$\left(\hat\rho_{Bl}\right)^2=\hat\rho_{Bl}$, $\hat\rho_{Al}$,
$\hat\rho_{Bl}$ are the density operators of subsystems. With the
validity of equality (4) the system is considered separable that
marked here by additional subscript $s$. In the separable system
quantum correlation (entanglement, quantum non-locality) is
absent. In this case statistical weights $p_{l}$ it is accepted to
call the hidden parameters of quantum system. The states of the
composite quantum system of form (4) can be obtained by local
operations (3) from the originally separated subsystems.

Answer to the question: is the state (the density operator) of
composite quantum system separable, or entangled? - it is accepted
to call the separability or disentanglement problem. At present
the mathematical disentanglement problem seems unsolvable in the
general case for two reasons \cite{batle}.  From one side there is
an infinite set of expansions of the separable density operator of
form (4), since the number of clean and not compulsorily
orthonormal states of subsystems is unconfined by anything.
Physically to this the infinite set of the ways of creating this
separable state with the aid of the local operations corresponds.
From other side there is an intriguing mathematical property: the
linear combination of the entangled states can be the separable
state. But at the same time the linear combination of the
separable states is always the separable state.

In what the physical sense of quantum correlation and why must be
known, are entangled subsystems or not? The traditional setting of
physical experiments consists of the observation of the dynamics
of physical subsystems as a result of their interaction. In many
cases the state of composite system, described by the density
operator, is obtained as a result of solving (most frequently the
approximate) dynamic Schroedinger (Neumann) equation as the result
of certain interaction between originally separated subsystems.
The mixed character of this state (i.e. its description by the
density matrix, but not by wave function) can appear, only if
there are some uncontrollable degrees of freedom out of the
subsystems in question. If state is described by the entangled
density matrix, then between the subsystems there is (or there
existed and it left its track) certain coherent interaction. But
if the density matrix is separable, then was unknown interaction
between the subsystems coherent, or not, and did remain there what
or tracks of the mutual coherence of subsystems. Here term
coherence is identical to term "quantum coherence", i.e., the
importance of the phase relationships between the subsystems in
the process of interaction and up to the moment of measurement.

Attempts at the mathematical solution of the disentanglement
(separability) problem compose separate direction in the quantum
informatics (in the linear algebra to it the unresolved problem of
the characterization of positive maps corresponds)
\cite{batle},\cite{terhal},\cite{plenio}. All approaches use as
the criterion of separability disturbance of one of the physical
limitations to the characteristics of the density operator. Let us
group the mathematical criteria of the separability (entanglement)
proposed in their physical content.

\begin{enumerate}

    \item \textbf{Positive partial transpose (PPT) criterion.}
    This is necessary, but insufficient criterion of separability.
    In order to clearly present its content let us write down the
    density operator $\hat\rho_{A+B}$ in the selected basis:
    \begin{equation}\label{eq5}
    \hat\rho_{(A+B)}=\sum_{\alpha,\alpha'}\sum_{\beta,\beta'}
    \langle\alpha\beta\vert\hat\rho_{(A+B)}\vert\alpha'\beta'\rangle
    \vert\alpha\rangle_{A}\langle\alpha'\vert\otimes
    \vert\beta\rangle_{B}\langle\beta'\vert.
    \end{equation}
    Then the density operator partially transposed on the subsystem
    $A$ is determined by the expression:
    \begin{equation}\label{eq6}
    \left(\hat\rho_{(A+B)}\right)_{T(A)}=\sum_{\alpha,\alpha'}
    \sum_{\beta,\beta'}
    \langle\alpha\beta\vert\hat\rho_{(A+B)}\vert\alpha'\beta'\rangle
    \vert\alpha'\rangle_{A}\langle\alpha\vert\otimes
    \vert\beta\rangle_{B}\langle\beta'\vert.
    \end{equation}
    System is separable, if this operator
    $\left(\hat\rho_{(A+B)}\right)_{T(A)}$ has only positive
    eigenvalues. (It is equivalent for the
    transposition on the subsystem $B$:
    $\left([\hat\rho_{(A+B)}\right)_{T(B)}$, moreover
    $\left(\left(\hat\rho_{(A+B)}\right)_{T(A)}\right)_{T(B)}=
    \left(\left(\hat\rho_{(A+B)}\right)_{T(B)}\right)_{T(A)}=
    \left(\hat\rho_{(A+B)}\right)_{T}$).
    The condition of positive partial transposition makes
    simple physical sense. The separable system satisfies
    condition (4), which shows that the dynamics of subsystems
    occurs it independently and, therefore, satisfies the
    condition of local unitarity (3), i.e., reversibility
    in the time. The operation of transposition on the subsystem
    corresponds to time reversal in this subsystem.
    Consequently, the condition of positivity corresponds to the
    absence of nonphysical (negative) probabilities with
    the time reversal in one of the subsystems.
    \item \textbf{Reduction criterion.}
    In the mathematical algorithm the reduction criterion is close
    to PPT criterion, since the conversion on one of the subsystems
    there also is done. It is possible to formulate this criterion
    as follows: system is in the separable state,
    if inequalities simultaneously are fulfilled \cite{batle}:
    \begin{equation}\label{eq7}
    \frac{1}{N_{A}}\hat I_{A}\otimes\hat\rho_{B\textrm{N}}-
    \hat\rho_{(A+B)}\ge 0,\,\,
    \hat\rho_{A\textrm{N}}\otimes\frac{1}{N_{B}}\hat I_{B}-
    \hat\rho_{(A+B)}\ge 0,
    \end{equation}
    where $\hat\rho_{A\textrm{N}}=\textrm{Sp}_{B}\hat\rho_{(A+B)}$,
    $\hat\rho_{B\textrm{N}}=\textrm{Sp}_{A}\hat\rho_{(A+B)}$ are
    reduced density operators of subsystems. Inequalities for
    the eigenvalues of operators here are implied,
    i.e. is assumed passage to the basis, in which the operators
    are diagonal.\\
\end{enumerate}

The enumerated criteria carry to the operational, i.e., to such,
for which the algorithm of calculation and relationship for some
values, which determine the conditions of separability, is
indicated. The physical sense of last criterion is different from
that indicated in the first point. It is possible to connect it
with distinctions in kind in the entropy of the entangled and
separeble composite quantum systems. Therefore let us transfer the
first known properties of the entropy of composite quantum
systems. Very determination of entropy is in this case debating
point with the ambiguous and sometimes contradictory
terminology.\\

The quantity of information is evaluated , which we know about the
quantum system globally (i.e. $\hat\rho_{(A+B)}$) and locally
(most frequently there is in the form $\hat\rho_{A\textrm{N}}$,
$\hat\rho_{B\textrm{N}}$). First remind of the von Neumann's
determination of the entropy of the quantum system, described by
the density operator $\hat\rho$:
\begin{equation}\label{eq8}
    S\left(\hat\rho\right)=-
    \textrm{Sp}\left(\hat\rho\ln\hat\rho\right).
    \end{equation}
It is assumed that with the calculation of this value they pass to
the basis, in which the density matrix is diagonal. Entropy
describes the deflection of quantum system from the pure state.
Its property:
\begin{itemize}
\item Entropy is equal to zero (it is minimum!) for the pure
state. Entropy is maximum and equal to $\ln N$ for a maximally
mixed state, when $\hat\rho=(1/N)\hat I$. In the first case is
known maximally possible, and in the second - minimally possible
information about the quantum system \cite{fano}. Let us note the
opposition of terminology from the accepted in the quantum
informatics definition of the characteristics of the states of
quantum system from the possibility of the content in them of
minimum ($S(\hat\rho)=0$) and maximum ($S(\hat\rho)=\max$)
quantity of information \cite{gorb}. Entropy is invariant with the
unitary conversions of the basis: $S(\hat\rho)=S\left(\hat
U\hat\rho\hat U^{\dagger}\right)$.\\
\item  The entropy of the separable quantum systems is additive,
i.e.\\
$S\left(\hat\rho_{A}\otimes\hat\rho_{B}\right)=
S(\hat\rho_{A})+S(\hat\rho_{B})$. However, for the composite
entangled quantum system is characteristic the property of
sub-additivity, determined by the inequality: $\left\vert
S(\hat\rho_{A\textrm{N}})-S(\hat\rho_{B\textrm{N}})\right\vert\le
S(\hat\rho_{A+B})\le
S(\hat\rho_{A\textrm{N}})+S(\hat\rho_{B\textrm{N}})$. According to
Shannon's theory, the entropy of composite system never can be
less than the entropy of any of its parts. For the entangled
system with the local reduced (according to von Neumann's
algorithm) density operators  - this is incorrect. It is
considered that this property of entropy can serve as the
indicator of the entanglement of state. However, there is a
separate direction in quantum informatics, which argues the
inapplicability of the classical determination of information to
the composite quantum systems and is proposed a number of the
alternative determinations of entropy for the composite quantum
systems for the purpose to return by it the property of additivity
\cite{batle}.
\end{itemize}

It is considered that existence of entanglement experimentally is
manifested in the inequality $\langle\hat A\hat
B\rangle\ne\langle\hat A\rangle_{\textrm{N}}\langle\hat
B\rangle_{\textrm{N}}$. To the left stands the experimentally
specific average value of that nonlocal observable, specific by
expression $\langle\hat A\hat
B\rangle=\textrm{Sp}\hat\rho_{A+B}\left(\hat A\otimes\hat
B\right)$. The product of the calculated local average values,
determined by equations $\langle\hat
A\rangle_{\textrm{N}}=\textrm{Sp}\hat\rho_{A+B}\hat A'$,
$\langle\hat B\rangle_{\textrm{N}}=\textrm{Sp}\hat\rho_{A+B}\hat
B'$ to the right stands. The upper prime noted operators, extended
to the complete space: $\hat A'=\hat A\otimes\hat I_{B}$, $\hat
P'_{B}(\beta)=\hat I_{A}\otimes\hat P_{B}(\beta)$. It is accepted
to name the specially selected nonlocal observable $\hat W$ for
the concrete systems the witness of entanglement. It is proven
that the composite bipartite system is entangled if and only if
there is witness of entanglement - Hermitian operator $\hat
W\,(\hat W=\hat W^{\dagger})$, for whom are valid the
inequalities: $\textrm{Sp}\hat\rho_{A+B}\hat W\le 0$ while
$\textrm{Sp}\hat\rho_{(A+B)s}\hat W\ge 0$ for all separable states
$\hat\rho_{(A+B)s}$. This is the sufficiently abstract
non-operational criterion of quantum correlation, to which it is
difficult to compare literal physical sense, however precisely
this form of the inequality (them proposed much for the different
experimental situations) is taken asd the experimental test of the
quantum correlation of subsystems.\\

Let us emphasize that it is always assumed that with the variety
of approaches to the description of quantum correlation the local
operators of the density of subsystems (and, correspondingly, the
local dynamics of subsystems) are determined by the reduction
algorithm of von Neumann
($\hat\rho_{A\textrm{N}}(t)=\textrm{Sp}_{B}\hat\rho_{A+B}(t)$,
$\hat\rho_{B\textrm{N}}(t)=\textrm{Sp}_{A}\hat\rho_{A+B}(t)$). Our
work is directed toward the assertion of idea about the
approximate nature of von Neumann's algorithm of the determination
of the local density operators, its injustice in the general case
and the needs for the calculation of the mutual correlation of
subsystems with the description of their local dynamics.\\

The idea of examination is based on what non-universality of von
Neumann's algorithm is already repeatedly demonstrated in
different physical tasks. Apparently, historically as this first
example can serve the construction of the quantum theory of
relaxation \cite{fano},\cite{blum}. There one of the interacting
subsystems (let $B$) from the physical considerations is relied
stationary, that possesses by the properties of quasi-classical
thermostat. Its state is approximately described by the
Boltzmann's density operator with the specific temperature
$\textrm{T}$: $\hat\rho_{B}(t)\approx\hat\rho_{B}(\textrm{T})=
\exp\left(-\left(1/k_{\textrm{B}}T\right)\hat H_{B}\right)$. For
the definition of the state of subsystem $A$ the approximate
disentanglement was postulated in the form of the relationship:
$\hat\rho_{A+B}(t)\approx\hat\rho^{d}_{A+B(\textrm{T})}(t)=
\hat\rho_{A}(t)\otimes\hat\rho_{B}(\textrm{T})$ , known as the
"first assumption of the quantum theory of relaxation" or the
"basic condition of irreversibility"
\cite{fano},\cite{blum},\cite{prigozh} (here and further by
superscript $d$ we note the approximate disentanglement).\\

Another approximate algorithm of disentanglement, widely utilized
in the quantum informatics, is based on the positions of the
quantum theory of measurements
\cite{preskill},\cite{zurek},\cite{men},\cite{koshino}, when both
subsystems are relied by quantum. It was initially postulated that
the indirect determination of the state of subsystem $A$ is
accomplished by means of the individual ideal projection quantum
measurement on quasi-independent subsystem-pointer $B$
\cite{zurek},\cite{koshino}. As a result of measurement of the
observable $\hat B$ by external macroscopic gauge the subsystem
$B$ (according to the projection postulate of quantum mechanics)
occurs in one of the eigenstates $\vert\beta\rangle$ of that
observable. The density operator of subsystem-pointer after
measurement is immediately considered equal:
$\hat\rho_{B(\beta)}(t)=\hat
P_{B}(\beta)=\vert\beta\rangle\langle\beta\vert$. The density
operator of subsystem $A$ in this case is defined as the result of
the quantum averaging of the density operator $\hat\rho_{A+B}(t)$
over the subsystem $B$ \cite{koshino}:
\begin{equation}\label{eq9}
    \hat\rho_{A(\beta)}(t)=\frac{\textrm{Sp}_{B}
    \left(\hat\rho_{A+B}(t)\hat P'_{B}(\beta)\right)}
    {\textrm{Sp}_{AB}\left(\hat\rho_{A+B}(t)\hat
    P'_{B}(\beta)\right)}.
    \end{equation}
Disentangled state:
\begin{equation}\label{eq10}
    \hat\rho_{A+B}(t)\to\hat\rho^{d}_{A+B(\beta)}=
    \hat\rho_{A(\beta)}(t)\otimes\hat P_{B}(\beta),
    \end{equation}
is considered as the initial state (relative to the moment of
measurement) in the description of the subsequent dynamics of
composite quantum system, which leads to the quantum Zeno effect
\cite{koshino}. Such type indirect projection measurements
(Zeno-like measurements) on the quasi-independent subsystem widely
are discussed as one of the possible mechanisms of control of the
state of subsystems in the quantum information theory.\\

In the recent works \cite{vedral},\cite{grisha} was shown the
importance of the calculation of the entanglement of quantum
subsystem with the external (with respect to the considered
composite system) gauge in the description of the result of
quantum measurement. Is noted, that by the consequence of this
calculation can be the incomplete loss of quantum coherence
subsystem, i.e., a difference in its local density operator after
measurement from the linear combination of projectors.\\

In the enumerated non-Neumann's algorithms of the approximate
disentanglement the desired state of the observed subsystem $A$ is
determined in the form of functional from the known density
operator of the of composite system $\hat\rho_{A+B}(t)$ and
density operator of subsystem $B$, postulated or obtained from the
additional physical considerations. With the use of a traditional
Neumann's algorithm mathematically equivalent situation occurs -
the state of composite system $\hat\rho_{A+B}(t)$ is known, and is
postulated the algorithm of the calculation of the reduced density
operator of subsystem $A$, $\hat\rho_{A\textrm{N}}(t)$. The
inverse problem is obvious: to what state of subsystem $B$,
$\hat\rho_{B?}(t)$ does correspond the Neumann's determination of
the density operator of the "observable" subsystem $A$? Or, to
what disentanglement
$\hat\rho_{A+B}(t)\to\hat\rho_{A\textrm{N}}\otimes\hat\rho_{B?}(t)$
it does correspond? Its examination was the starting point of ours
work.\\

 \section{Physical content of the operation of the
 reduction-disentanglement of von Neumann.}

 The necessary step of the approximate disentanglement is answer
 to the question: should be represented the density operator of
 composite system in the form of the tensor product of the
 local density operators of subsystems (tensor product structure
 \cite{zanardi}), or in the form (4) of the linear combination
 of such products? In the quantum information theory usually
 the first idea is postulated \cite{preskill}. However, in the
 recent work \cite{zanardi} the arguments were formulated,
 which show that precisely this selection is dictated by
 the experimental determination of observables and
 interactions between the subsystems. The idea of work
 \cite{zanardi} lies in the fact that the determination of
 composite system includes the possibility of conducting of
 local operations and measurements. Such composite system
 from an experimental point of view appears as two
 quasi-independent correlated subsystems, to what its
 mathematical idea in the form of the tensor product of
 the local density operators of subsystems corresponds.
 Assuming the positions of work \cite{zanardi}, we will
 represent the entangled density operator in the form of
 the tensor product of some correlated density operators
 of subsystems:
 \begin{equation}\label{eq11}
    \hat\rho_{A+B}(t)\approx\hat\rho^{d}_{A+B)}=
    \hat\rho_{Ac}(t)\otimes\hat\rho_{Bc}(t),
    \end{equation}
where subscript $c$ distinguishes the local correlated density
operators from corresponding local reduced  density operators.\\

We transform (11) by the method, analogous to the method, used by
von Neumann for the proof of the mutual correlation of the average
values of observables (see chapter 6.2 \cite{von}). Let us
multiply (11) to the right on $\hat\rho'_{Bc}(t)$ or
$\hat\rho'_{Ac}(t)$ and let us take partial track on the subsystem
$B$ or $A$. Using the orthonormality of the correlated density
operators of subsystems, we will obtain the system of two
connected equations \cite{solyear},\cite{soljetpl},\cite{solalg}:
\begin{equation}\label{eq12}
    \hat\rho_{Ac}(t)\approx\frac{\textrm{Sp}_{B}\hat\rho_{A+B}(t)
    \hat\rho'_{Bc}(t)}
    {\textrm{Sp}_{AB}\hat\rho_{A+B}(t)
    \hat\rho'_{Bc}(t)},\quad
    \hat\rho_{Bc}(t)\approx\frac{\textrm{Sp}_{A}\hat\rho_{A+B}(t)
    \hat\rho'_{Ac}(t)}
    {\textrm{Sp}_{AB}\hat\rho_{A+B}(t)
    \hat\rho'_{Ac}(t)}.
    \end{equation}
Right sides of (12) are the normalized quantum averaging of the
density operator of the closed system over one of subsystems. Each
equation defines the density operator of one subsystem as
functional from the density operator of the closed system and
density operator of another subsystem. Actually, this simply the
more convenient record of expression (11). Such relationships are
not unique. Repeating the procedure of right multiplication it is
possible to obtain the set of the expressions of the form:
\begin{equation}\label{eq13}
    \hat\rho_{Ac}(t)\approx\frac{\textrm{Sp}_{B}\hat\rho_{A+B}(t)
    \left[\hat\rho'_{Bc}(t)\right]^{m}}
    {\textrm{Sp}_{AB}\hat\rho_{A+B}(t)
    \left[\hat\rho'_{Bc}(t)\right]^{m}},\quad
    \hat\rho_{Bc}(t)\approx\frac{\textrm{Sp}_{A}\hat\rho_{A+B}(t)
    \left[\hat\rho'_{Ac}(t)\right]^{m}}
    {\textrm{Sp}_{AB}\hat\rho_{A+B}(t)
    \left[\hat\rho'_{Ac}(t)\right]^{m}},
    \end{equation}
where $m$ - arbitrary integer. It is obvious that essential it is
possible to consider only expressions with
$m\le\left(N^{2}_{A(B)}-1\right)$ for the linearly independent
degrees of the correlated density operators of subsystems.\\

Now let us show, to what disentanglement does implicitly
correspond von Neumann's reduction? Let us examine the first of
expressions (13), which determine the state of the "observed"
subsystem $A$. Assume that its correlated density operator is
determined by von Neumann's reduction
$\hat\rho_{Ac}(t)=\hat\rho_{A\textrm{N}}(t)$. Then precise
equalities for all expressions (13) are carried out in two
cases:\\
\begin{enumerate}
\item Or the density operator of subsystem $B$ corresponds to the
pure state, when
$\left[\hat\rho_{Bc}(t)\right]^{m}=\hat\rho_{Bc}(t)$ (that
corresponds physically to the limiting case of the noninteracting
subsystems).\\
\item Or it is proportional to the unit operator
$\hat\rho_{Bc}(t)=\hat\rho_{B\max}(t)=(1/N_{B})\hat I_{B}$ (that
corresponds to the steady state of subsystem $B$ with the maximum
entropy, the minimum information about subsystem or the infinite
temperature according to \cite{fano}).
\end{enumerate}
Since the trivial case of the noninteracting subsystems does not
correspond to initial assumption about the entanglement of
composite quantum system, remains the second version. Thus, the
use of von Neumann's algorithm of the reduction for determining
the local density operator of the observed subsystem is physically
equivalent to the approximate disentanglement, with which the
unobservable subsystem is relied by being stationarily been in the
state with the maximum entropy or infinite temperature, in which
it with the equal probability is found in any from the
eigenstates. In this case quantum coherence (non-diagonal matrix
elements of density matrix) in the subsystem $B$ is equal to zero:
\begin{equation}\label{eq14}
    \hat\rho_{A+B}(t)\to\hat\rho^{d}_{A+B}(t)=
    \hat\rho_{A\textrm{N}}(t)\otimes\hat\rho_{B\max}.
    \end{equation}
To the same conclusion it is possible to come by another way,
using positions of the quantum measurements theory
\cite{zurek},\cite{men},\cite{koshino}. Let the subsystem $B$ be
the quantum pointer, with the aid of which is accomplished the
indirect measurement of the state of subsystem $A$. What must be
the results of many projective measurements (1),(2) on the equally
prepared system $\hat\rho_{A+B}(t)$, so that the result would be
described by the density operator of
$\hat\rho_{A\textrm{N}}(t)$?\\

Let $p(\beta)$ is the probability of observing the value $\beta$,
which is appeared the eigenvalue of that observable $\hat B$, with
the individual ideal projection quantum measurements on the
quasi-independent subsystem $B$ \cite{koshino}. With each
measurement the density operator of subsystem $A$ is determined by
the appropriate expression of form (2). Consequently, takeing into
account the statistical nature of $\beta$ measurements, the
density operator of subsystem $A$ it is necessary to define as the
result of statistical averaging over the results of many
individual projection measurements of $\hat B$:
\begin{equation}\label{eq15}
    \hat\rho_{AP(B)}(t)=\frac{\sum_{\beta}p(\beta)\textrm{Sp}_{B}
    \left(\hat\rho_{A+B}\hat P'_{B}(\beta)\right)}
    {\textrm{Sp}_{A}\sum_{\beta}p(\beta)\textrm{Sp}_{B}
    \left(\hat\rho_{A+B}\hat P'_{B}(\beta)\right)}.
    \end{equation}
Using permutability of the operations of summation over $\beta$
and the takings of the track on the subsystem $B$, let us write
down the condition of the equality of this averaged density
operator of subsystem $A$ to the operator
$\hat\rho_{A\textrm{N}}(t)$:
\begin{equation}\label{eq16}
    \frac{\textrm{Sp}_{B}\left(\hat\rho_{A+B}(t)\left(\hat
    I_{A}\otimes\sum_{\beta}p(\beta)\hat P_{B}(\beta)\right)
    \right)}
    {\textrm{Sp}_{AB}\left(\hat\rho_{A+B}(t)\left(\hat
    I_{A}\otimes\sum_{\beta}p(\beta)\hat P_{B}(\beta)\right)
    \right)}=\textrm{Sp}\hat\rho_{A+B}(t).
    \end{equation}
Equality (16) is correct in the general case of arbitrary state
only if $p(\beta)=\left(1/N_{\beta}\right)$. Thus, the description
of the dynamics of the observed subsystem by the von Neumann's
reduced density operator corresponds in the quantum measurements
theory to the case, when quantum subsystem-pointer with the
projection measurements with the equal probability is revealed in
any of the eigenstates. The use of a projection postulate in the
quantum measurements theory determines the total loss of quantum
coherence by subsystem-pointer with each individual measurement
\cite{koshino},\cite{dass}. Comparing (16) and (9) we conclude
that the state of subsystem-pointer in this case is physically
equivalent to its presence in the state with the infinite
temperature. By other words the use of von Neumann reduction with
the disentanglement corresponds precisely to the "first assumption
of the quantum relaxation theory" in the extreme case, when the
temperature of thermostat is relied infinite
\cite{fano},\cite{blum}.\\

If we use an idea about the correlation of subsystems (12)-(13),
then von Neumann's reduction \cite{von}, the "first assumption of
the quantum theory of relaxation" \cite{fano},\cite{blum} and
definition (9) \cite{koshino} can be examined as special cases of
approaching the assigned state of the "unobservable" subsystem in
the disentanglement problem. In each case in the explicit or
implicit form the state of one subsystem is postulated, and the
state (density operator) of another subsystem correlated with it
is determined by expressions (13). It is possible to conclude that
in the conventional algorithm of the calculation of the average
values of those observed by one of the subsystems the measuring
projection postulate of quantum mechanics is implicitly used
twice. For the first time with the determination of the density
operator of subsystem from the known state of composite system
(reduction-disentanglement of von Neumann), and for the second
time with the calculation of the average values of the observables
for the subsystem with the counted already known density operator.
Our further consideration is based on the idea, that the first
step (disentanglement) does not identify with the macroscopic
projection measurement, and it must be based on the basis of the
conditions of each specific objectives.\\

\section{Correlated disentanglement.}

The algorithms of the approximate disentanglement examined include
in the explicit or implicit form assumption about state of one of
the subsystems. By each concrete selection of the density operator
of the "unobservable" subsystem is simulated the specific physical
process of measurement and opposite effect of measurement on the
state of subsystems. This idea for the projection measurements is
contained in the quantum theory of measurements
\cite{zurek},\cite{men},\cite{koshino} in the form of the
mathematical expressions of form (9), (15). For the clarity let us
show differences in the calculated dynamics of the observed
subsystem, caused by the selection of the operator of the density
of the unobservable subsystem, based on the example to model in
the quantum informats of the composite system of two qubits
\cite{preskill},\cite{dass}.

Let us designate the eigenstates of the independent
subsystems-qubits $A$ and $B$, $\vert 2\rangle_{A},\vert
1\rangle_{A}$ and $\vert 2\rangle_{B},\vert 1\rangle_{B}$,
respectively. The operators of composite system are represented by
the matrices of the fourth order, whose each element is designated
by two pairs of the subscripts $(\alpha\beta,\alpha'\beta')$ (see,
for example, \cite{blum} Appendix A). The first index of each pair
$(\alpha,\alpha')$ corresponds to the eigenstate of subsystem $A$,
and the second  $\beta,\beta'$ - subsystem $B$. In the general
case the unitary dynamics of the closed system in question is
described by the density matrix  of form \cite{blum}:
\begin{equation}\label{eq17}
   \left(\hat\rho_{A+B}(t)\right)=\left(
   \begin{array}{cccc}
   \rho_{22,22} & \rho_{22,21} & \rho_{22,12} & \rho_{22,11}\\
   \rho_{21,22} & \rho_{21,21} & \rho_{21,12} & \rho_{21,11}\\
   \rho_{12,22} & \rho_{12,21} & \rho_{12,12} & \rho_{12,11}\\
   \rho_{11,22} & \rho_{11,21} & \rho_{11,12} & \rho_{11,11}\\
   \end{array}
   \right),
    \end{equation}
where dependence on the time is omitted for simplicity in the
expressions of matrix elements.\\

According to (15) the dynamics of qubit $A$, determined from the
results of many projective measurements of qubit $B$ with the
measured probabilities $p_{B2},p_{B1}$, to reveal it in the states
$\vert 2\rangle_{B},\vert 1\rangle_{B}$, correspondingly, is
described by the density operator:
\begin{equation}\label{18}
\hat\rho_{AP(B)}=\frac{p_{B2}\textrm{Sp}_{B}
\left(\hat\rho_{A+B}\left(\hat I_{A}\otimes\hat
P_{B2}\right)\right)+p_{B1}\textrm{Sp}_{B}
\left(\hat\rho_{A+B}\left(\hat I_{A}\otimes\hat
P_{B1}\right)\right)} {\textrm{Sp}_{A}\left( p_{B2}\textrm{Sp}_{B}
\left(\hat\rho_{A+B}\left(\hat I_{A}\otimes\hat
P_{B2}\right)\right)+p_{B1}\textrm{Sp}_{B}
\left(\hat\rho_{A+B}\left(\hat I_{A}\otimes\hat
P_{B1}\right)\right)\right)}.
\end{equation}
Let $p_{B2}=p,\, p_{B1}=1-p$, i.e., the density matrix of the
subsystem of qubit-pointer be relied by stationary and equal:
\begin{equation}\label{eq19}
\left(\hat\rho_{B(P)}\right)= \left(
\begin{array}{cc}
 p & 0\\
 0 & 1-p
 \end{array}
 \right).
\end{equation}
Then the local, correlated with it density matrix, which describes
the dynamics of the observed subsystem $A$, is equal:
\begin{multline}\label{eq20}
\left(\hat\rho_{AP(B)}\right)= \left(\begin{array}{cc}
p\rho_{22,22}+(1-p)\rho_{21,21} &
p\rho_{22,12}+(1-p)\rho_{21,11}\\
p\rho_{12,22}+(1-p)\rho_{11,21} & p\rho_{12,12}+(1-p)\rho_{11,11}
\end{array}\right)\times\\
\times\left[p\rho_{22,22}+(1-p)\rho_{21,21}+
p\rho_{12,12}+(1-p)\rho_{11,11}\right]^{-1}.
\end{multline}
This result immediately is obtained, if we substitute (19) into
the first of (12). Thus, the approximate disentanglement is
realized in this case by the following replacement:
\begin{multline}\label{eq21}
\left(\hat\rho_{A+B}\right)\to\left(\hat\rho^{d}_{A+B}\right)=
\left(\hat\rho_{AP(B)}\right)\otimes\left(\hat\rho_{B(P)}\right)=
\\
\left(\begin{array}{cc}
p^2\rho_{22,22}+p(1-p)\rho_{21,21} & 0 \\
0 & p(1-p)\rho_{22,22}+(1-p)^{2}\rho_{21,21}\\
p^2\rho_{12,22}+p(1-p)\rho_{11,21} & 0 \\
0 & p(1-p)\rho_{12,22}+(1-p)^{2}\rho_{21,21}
\end{array}\right.\\
\left.\begin{array}{cc} p^2\rho_{22,12}+p(1-p)\rho_{21,11} & 0\\
0 & p(1-p)\rho_{22,12}+(1-p)^{2}\rho_{21,11}\\
p^2\rho_{12,12}+p(1-p)\rho_{11,11} & 0\\
0 & p(1-p)\rho_{12,12}+(1-p)^{2}\rho_{11,11}
\end{array}
\right)\times\\
\left[p\rho_{22,22}+(1-p)\rho_{21,21}+p\rho_{12,12}+(1-p)\rho_{11,11}
\right]^{-1}
\end{multline}
This nontraditional presentation of the operation of reduction,
recorded in the "disentangled" form of the direct product of the
local density operators of subsystems, makes it possible to
clearly present the sense of the made approximations, comparing
matrix elements (21) and (17). The first difference - the
non-diagonal on the indices of subsystem-pointer matrix elements
of the density matrix of composite system are assumed equal to
zero. By this step is mathematically reflected the postulated loss
by the subsystem-pointer of quantum coherence (decoherence) with
the external macroscopic projection measurement on the subsystem
\cite{zurek},\cite{dass}. Let us note that in this case the
quantum coherence in the subsystem $A$ does not disappear, i.e.,
complete dekogerence of both the composite system and the observed
subsystem does not occur. The second difference - the diagonal on
the indices of subsystem-pointer elements of complete density
matrix are substituted with their linear combinations with the
weights, determined according to the results of local projection
measurements on the subsystem-pointer $B$. With the equal
probability of detecting qubit $B$ in the eigenstates ($p=1/2$)
the approximate disentanglement corresponds to the von Neumann's
reduction. With $p\ne 1/2$ joint use of both mathematical
approximations is equivalent to physical assumption about the
validity of the description of the state of the unobservable
subsystem by the specific temperature $\textrm{T}$. It is evident
that with $p\ne 1/2$ the matrix elements of the local density
operator (20) quantitatively are differed from that obtained by
traditional calculation and, correspondingly, the dynamics of the
observed subsystem can differ from usually supposed.\\

Qualitative difference in the dynamics of the observed subsystem
$A$ will arise, if one assumes that the quantum coherence remains
with the disentanglement in the subsystem $B$. To this assumption
corresponds idea about that which the physical separation of
composite system into the subsystems does not identify with
conducting of projection macroscopic measurement. Either the
physical separation of subsystems is accomplished before
conducting of local projection measurement on qubit $B$, or the
nondestructive quantum measurement is conducted . To the retention
of quantum coherence in the subsystem $B$ mathematically
corresponds adding to the nondiagonal elements of the local
density matrix $\hat\rho_{B}$ of values $\vert b\vert\ne 0$
instead of their equality to zero, postulated in (19):
\begin{equation}\label{eq22}
\left(\hat\rho_{Bc}\right)=\left(\begin{array}{cc} p & b\\
b^{*} & 1-p
\end{array}\right),\quad
\left(\hat\rho'_{Bc}\right)=\left(\hat I_{A}\right)\otimes
\left(\hat\rho_{Bc}\right)=\left(\begin{array}{cccc} p & b & 0 & 0\\
b^{*} & 1-p & 0 & 0\\
0 & 0 & p & b\\
0 & 0 & b^{*} & 1-p
\end{array}\right).
\end{equation}
Using first of (12) we will obtain correlated to (22) the local
density matrix of the subsystem $A$:
\begin{multline}\label{eq23}
\left(\hat\rho_{Ac}\right)=\\
=\left(\begin{array}{cc} \begin{array}{c}
p\rho_{22,22}+b^{*}\rho_{22,21}+\\
+b\rho_{21,22}+(1-p)\rho_{21,21}\end{array} & \begin{array}{c}
p\rho_{22,12}+b^{*}\rho_{22,11}+\\
+b\rho_{21,12}+(1-p)\rho_{21,11}\end{array}\\ \begin{array}{c}
p\rho_{12,22}+b^{*}\rho_{12,21}+\\
+b\rho_{11,22}+(1-p)\rho_{11,21}\end{array} & \begin{array}{c}
p\rho_{12,12}+b^{*}\rho_{12,11}+\\
+b\rho_{11,12}+(1-p)\rho_{11,11}\end{array}
\end{array}\right)\times\\
\times\left[p\left(\rho_{22,22}+\rho_{12,12}\right)+
(1-p)(\left(\rho_{21,21}+\rho_{11,11}\right)+\right.\\
+\left.b\left(\rho_{21,22}+\rho_{11,12}\right)+
b^{*}\left(\rho_{22,21}+\rho_{12,11}\right)\right]^{-1}
\end{multline}
From comparison (23) with (20) it is evident that the dynamics of
populations (diagonal elements $\left(\hat\rho_{Ac}\right)$) and
quantum coherence of qubit $A$, depends in this case not only on
probabilities to reveal qubit $B$ in the eigenstates, but also
from its quantum coherence. The analogous (21) expression of the
approximately disentangled density matrix of composite system in
this case is equal to:
\begin{equation}\label{eq24}
\begin{array}{c}
\left(\hat\rho_{A+B}\right)\to\left(\hat\rho^{d}_{(A+B)c}\right)=
\left(\hat\rho_{A(Bc)}\right)\otimes\left(\hat\rho_{Bc)}\right)=
\\
\left(\begin{array}{cc}
\begin{array}{c}
p^2\rho_{22,22}+p(1-p)\rho_{21,21}+\\
pb^{*}\rho_{22,21}+ pb\rho_{21,22}
\end{array} &
\begin{array}{c}
pb\rho_{22,22}+(1-p)b\rho_{21,21}+\\
\vert b\vert^{2}\rho_{22,21}+ b^{2}\rho_{21,22}
\end{array}
\\
\begin{array}{c}
pb^{*}\rho_{22,22}+(1-p)b^{*}\rho_{21,21}+\\
\left(b^{*}\right)^{2}\rho_{22,21}+ \vert b\vert^{2}\rho_{21,22}
\end{array} &
\begin{array}{c}
p(1-p)\rho_{22,22}+(1-p)^{2}\rho_{21,21}+\\
(1-p)b^{*}\rho_{22,21}+ (1-p)b\rho_{21,22}
\end{array}
\\
\begin{array}{c}
p^2\rho_{12,22}+p(1-p)\rho_{11,21}+\\
pb^{*}\rho_{12,21}+pb\rho_{11,22}
\end{array} &
\begin{array}{c}
pb\rho_{12,22}+(1-p)b\rho_{11,21}+\\
\vert b\vert^{2}\rho_{12,21}+b^{2}\rho_{11,22}
\end{array}
\\
\begin{array}{c}
pb^{*}\rho_{12,22}+(1-p)b^{*}\rho_{11,21}+\\
\left(b^{*}\right)^{2}\rho_{12,21}+\vert b\vert^{2}\rho_{11,22}
\end{array} &
\begin{array}{c}
p(1-p)\rho_{12,22}+(1-p)^{2}\rho_{11,21}+\\
(1-p)b{*}\rho_{12,21}+(1-p)b\rho_{11,22}
\end{array}
\end{array}\right.
\\
\left.\begin{array}{cc}
\begin{array}{c}
p^2\rho_{22,12}+p(1-p)\rho_{21,11}+\\
pb^{*}\rho_{22,11}+ pb\rho_{21,12}
\end{array} &
\begin{array}{c}
pb\rho_{22,12}+(1-p)b\rho_{21,11}+\\
\vert b\vert^{2}\rho_{22,11}+ b^{2}\rho_{21,12}
\end{array}\\
\begin{array}{c}
pb^{*}\rho_{22,12}+(1-p)b^{*}\rho_{21,11}+\\
\left(b^{*}\right)^{2}\rho_{22,11}+ \vert b\vert^{2}\rho_{21,12}
\end{array} &
\begin{array}{c}
p(1-p)\rho_{22,12}+(1-p)^{2}\rho_{21,11}+\\
(1-p)b^{*}\rho_{22,11}+ (1-p)b\rho_{21,12}
\end{array}\\
\begin{array}{c}
p^2\rho_{12,12}+p(1-p)\rho_{11,11}+\\
pb^{*}\rho_{12,11}+pb\rho_{11,12}
\end{array} &
\begin{array}{c}
pb\rho_{12,12}+(1-p)b\rho_{11,11}+\\
\vert b\vert^{2}\rho_{12,11}+b^{2}\rho_{11,22}
\end{array}\\
\begin{array}{c}
pb^{*}\rho_{12,12}+(1-p)b^{*}\rho_{11,11}+\\
\left(b^{*}\right)^{2}\rho_{12,11}+\vert b\vert^{2}\rho_{11,12}
\end{array} &
\begin{array}{c}
p(1-p)\rho_{12,12}+(1-p)^{2}\rho_{11,11}+\\
(1-p)b{*}\rho_{12,11}+(1-p)b\rho_{11,12}
\end{array}
\end{array}
\right)\times
\\
\left[p\left(\rho_{22,22}+\rho_{12,12}\right)+
(1-p)\left(\rho_{21,21}+\rho_{11,11}\right)+\right.
\\
\left.b\left(\rho_{21,22}+\rho_{11,12}\right)+
b^{*}\left(\rho_{22,21}+\rho_{12,11}\right)\right]^{-1}.
\end{array}
\end{equation}
From comparison (21),(24) and respectively (20),(23) it is evident
that to idea about the retention of quantum coherence in the
subsystem-pointer qualitatively corresponds the partial retention
of the quantum coherence of composite system. In the general case
all nondiagonal elements of density matrix (24) are not equal to
zero. Simultaneously it is evident from this simplest example that
the use of approximation of the assigned state of the
subsystem-pointer (i.e. the selection of the model of measurement,
including as a special case, the conventional algorithm of the
reduction of von Neumann) leads always to redefining (change) of
all matrix elements of the composite density matrix. Thus always
with the realization of the mathematical operation of the
approximate disentanglement occurs a change in the entropy or
information about the composite quantum system \cite{vedral} and
partial decogerence \cite{dass}.\\

Different dynamics of the observed subsystem is the consequence of
different selection of the algorithm of disentanglement (model of
measurement).  The question about the criterion of the selection
of the disentanglement algorithm arises and its correspondence to
the experimental procedure of measurement accepted. Physically
limiting cases are the von Neumann's disentanglement, which
corresponds to complete decogerence of the subsystem-pointer and
the "nondestructive" disentanglement, to which corresponds the
invariability (in the limit) of the density operator of composite
system. If the approximation of the closed system and
nondestructive quantum measurement is correct, it is justified to
search for the algorithm of separation, which minimizes change of
$\hat\rho_{A+B}$ and simultaneously the change of entropy. Such
local mutually correlated density operators of subsystems it is
possible to obtain by solving system of equations (12) by the
method of sequential approximations \cite{solalg},\cite{soljetpl}:
\begin{multline}\label{eq25}
\hat\rho_{Ac}=\lim_{n\to\infty}\hat\rho^{(n+1)}_{A}=
\lim_{n\to\infty}\frac{\textrm{Sp}_{B}\hat\rho_{A+B}\hat\rho'^{(n)}_{B}}
{\textrm{Sp}_{AB}\hat\rho_{A+B}\hat\rho'^{(n)}_{B}},\\
\hat\rho_{Bc}=\lim_{n\to\infty}\hat\rho^{(n+1)}_{B}=
\lim_{n\to\infty}\frac{\textrm{Sp}_{A}\hat\rho_{A+B}\hat\rho'^{(n)}_{A}}
{\textrm{Sp}_{AB}\hat\rho_{A+B}\hat\rho'^{(n)}_{B}},
\end{multline}
where, for example, the reduced density operators
$\hat\rho_{A\textrm{N}},\hat\rho_{B\textrm{N}}$ it is possible to
use as zero approximation (n=0). One should emphasize that this
correlated reduction corresponds to idea about the equivalence of
the subsystems of the closed quantum system.

 \end{document}